\documentclass[submission,copyright,creativecommons]{eptcs}
\providecommand{\event}{DCM 2011} 
\usepackage{amsmath,amsfonts,amssymb,cite,graphicx,amsthm,enumerate,xypic}

\DeclareMathOperator{\Hom}{Hom}

\newcommand{\vecify}{{\mathcal V}}
\newcommand{\Act}{{A}}
\newcommand{\act}{{a}}
\newcommand{\Sit}{{S}}
\newcommand{\occ}{{v}}
\newcommand{\univ}{{\mathbf D}}
\newcommand{\uout}{{d_{out}}}
\newcommand{\uin}{{d_{in}}}
\newcommand{\mangle}{{\mathbf C}}

\newcommand{\psheaf}{{\mathcal F}}
\newcommand{\scat}{{\mathtt{Stoch}}}
\newcommand{\subs}{{\mathtt{Sys}}}
\newcommand{\mcat}{{\mathtt{Meas}}}
\newcommand{\eop}{{$\blacksquare$}}

\newcommand{\bra}{{\langle}}
\newcommand{\ket}{{\rangle}}

\newcommand{\bR}{{\mathbb R}}
\newcommand{\fm}{{\mathfrak m}}
\newcommand{\cP}{{\mathcal P}}

\newtheorem{thm}{Theorem}
\newtheorem{prop}[thm]{Proposition}
\newtheorem{cor}[thm]{Corollary}

\theoremstyle{remark}
\newtheorem{eg}{Example}
\newtheorem{rem}{Remark}
\newtheorem{defn}{Definition}

\title{On the information-theoretic structure of distributed measurements}
\author{David Balduzzi%
\footnote{I thank Dominik Janzing, Sanjeevi Krishnan and the reviewers for useful comments.}
\institute{Max Planck Institute for Intelligent Systems, T{\"u}bingen, Germany}
\email{david.balduzzi@tuebingen.mpg.de}}
\def\titlerunning{Structure of distributed measurements}
\def\authorrunning{David Balduzzi}
\begin{document}
\maketitle

\begin{abstract}
	The internal structure of a measuring device, which depends on what its components are and how they are organized, determines how it categorizes its inputs. This paper presents a geometric approach to studying the internal structure of measurements performed by distributed systems such as probabilistic cellular automata.  It constructs the quale, a family of sections of a suitably defined presheaf, whose elements correspond to the measurements performed by all subsystems of a distributed system. Using the quale we quantify (i) the information generated by a measurement; (ii) the extent to which a measurement is context-dependent; and (iii) whether a measurement is decomposable into independent submeasurements, which turns out to be equivalent to context-dependence. Finally, we show that only indecomposable measurements are more informative than the sum of their submeasurements.
\end{abstract}

\section{Introduction}
\label{s:intro}

Any classical physical system (by which we simply mean any deterministic function) can be taken as a measuring apparatus or input/output device. For example, a thermometer takes inputs from the atmosphere and outputs numbers on a digital display. The thermometer categorizes inputs by temperature and is blind to, say, differences in air pressure. 

Classical measurements are formalized as follows:
\begin{defn}
	\label{d:cmeasure}
	Given a classical physical system with state space $X$, a \emph{measuring device} is a function $f:X\rightarrow \bR$. The output $r\in\bR$ is the \emph{reading} and the pre-image $f^{-1}(r)\subset X$ is the \emph{measurement}.
\end{defn}

From this point of view a thermometer and a barometer are two functions, $T: X\rightarrow \bR$ and $B: X\rightarrow \bR$, mapping the state space $X$ of configurations (positions and momenta) of atmospheric particles to real numbers. When the thermometer outputs $2^\circ$, it specifies that the atmospheric configuration was in the pre-image $T^{-1}(2^\circ)$ which, assuming the thermometer perfectly measures temperature, is \emph{exactly} characterized as atmospheric configurations with temperature $2^\circ$. Similarly, the pre-images generated by the barometer group atmospheric configurations by pressure.

The classical definition of measurement takes a thermometer as a monolithic object described by a single function from atmospheric configurations to real numbers. The internal structure of the thermometer -- that is composed of countless atoms and molecules arranged in an extremely specific manner -- is swept under the carpet (or, rather, into the function).

This paper investigates the structure of measurements performed by \emph{distributed} systems. We do so by adapting Definition~\ref{d:cmeasure} to a large class of systems that contains networks of Boolean functions \cite{Kauffman:2003fv}, Conway's game of life \cite{gardner:70, berlekamp:82} and Hopfield networks \cite{hopfield:82, amit:89} as special cases.

Our motivation comes from prior work investigating information processing in discrete neural networks \cite{bt:08, bt:09}. The brain $X$ can be thought of as an enormously complicated measuring device $S\times X\xrightarrow{f} X$ mapping sensory states $s\in S$ and prior brain states $x\in X$ to subsequent brain states. Analyzing the functional dependencies implicit in cortical computations reduces to analyzing how the measurements performed by the brain are composed out of submeasurements by subdevices such as individual neurons and neuronal assemblies. The cortex is of particular interest since it seemingly effortlessly integrates diverse contextual data into a unified gestalt that determines behavior. The measurements performed by different neurons appear to interact in such a way that they generate more information jointly than separately. To improve our understanding of how the cortex integrates information we need to a formal language for analyzing how context affects measurements in distributed systems. 

As a first step in this direction, we develop methods for analyzing the geometry of measurements performed by functions with overlapping domains. We propose, roughly speaking, to study context-dependence in terms of the geometry of intersecting pre-images. However, since we wish to work with both probabilistic and deterministic systems, things are a bit more complicated.

We sketch the contents of the paper. Section \S\ref{s:stochastic} lays the groundwork by introducing the category of stochastic maps $\scat$. Our goal is to study finite set valued functions and conditional probability distributions on finite sets. However, rather than work with sets, functions and conditional distributions, we prefer to study stochastic maps (Markov matrices) between function spaces on sets. We therefore introduce the faithful functor $\vecify$ taking functions on sets to Markov matrices:
\begin{equation*}
	\Big[f:X\rightarrow Y\Big]\mapsto\Big[\vecify f:\vecify X\rightarrow \vecify Y\Big],
\end{equation*}
where $\vecify X$ is functions from $X$ to $\bR$. Conditional probability distributions $p(y|x)$ can also be represented using stochastic maps. 

Working with linear operators instead of set-valued functions is convenient for two reasons. First, it unifies the deterministic and probabilistic cases in a single language. Second, the dual $T^\natural$ of a stochastic map $T$ provides a symmetric treatment of functions and their corresponding inverse image functions. Recall the inverse of function $f:X\rightarrow Y$ is $f^{-1}:Y\rightarrow \underline{2}^X$, which takes values in the \emph{powerset} of $X$, rather than $X$ itself. Dualizing a stochastic map flips the domain and range of the original map, without introducing any new objects:
\begin{equation}
	\Big[f^{-1}:Y\rightarrow \underline{2}^X\Big]\,\mbox{ corresponds to }\,
	\Big[(\vecify f)^\natural:\vecify Y\rightarrow \vecify X\Big],
	\label{e:preimage-corr}
\end{equation}
see Proposition~\ref{t:preimage}.

Section \S\ref{s:fds} introduces distributed dynamical systems. These extend probabilistic cellular automata by replacing cells (space coordinates) with occasions (spacetime coordinates: cell $k$ at time $t$). Inspired by \cite{hooft:99, abramsky:09},  we treat distributed systems as collections of stochastic maps between function spaces so that processes (stochastic maps) take center stage, rather than their outputs. 
Although the setting is abstract, it has the advantage that it is \emph{scalable}: using a coarse-graining procedure introduced in \cite{balduzzi:11} we can analyze distributed systems at any spatiotemporal granularity.

Distributed dynamical systems provide a rich class of toy universes. However, since these toy universes do not contain conscious observers we confront Bell's problem  \cite{bell:90}: ``What exactly qualifies some physical [system] to play the role of `measurer'?'' In our setting, where we do not have to worry about collapsing wave-functions or the distinction between macroscopic and microscopic processes, the solution is simple: \emph{every} physical system plays the role of measurer. More precisely, we track measurers via the category $\subs_\univ$ of subsystems of $\univ$. Each subsystem $\mangle$ is equipped with a mechanism $\fm_\mangle$ which is constructed by gluing together the mechanisms of the occasions in $\mangle$ and averaging over extrinsic noise.

Measuring devices are typically analyzed by varying their inputs and observing the effect on their outputs. By contrast this paper fixes the output and \emph{varies the device over all its subdevices} to obtain a family of submeasurements parametrized by all subsystems in $\subs_\univ$. The internal structure of the measurement performed by $\univ$ is then studied by comparing submeasurements.

We keep track of submeasurements by observing that they are sections of a suitably defined presheaf. Sheaf theory provides a powerful machinery for analyzing relationships between objects and subobjects \cite{maclane:92}, which we adapt to our setting by introducing the structure presheaf $\psheaf$, a contravariant functor from  $\subs_\univ$ to the category of measuring devices $\mcat_\univ$ on $\univ$. Importantly, $\psheaf$ is \emph{not} a sheaf: although the gluing axiom holds, uniqueness fails, see Theorem~\ref{t:presheaf}. This is because the restriction operator in $\mcat$ is (essentially) marginalization, and of course there are infinitely many joint distributions $p(x,y)$ that yield marginals $p(x)$ and $p(y)$.

Section \S\ref{s:measurement} adapts Definition~\ref{d:cmeasure} to distributed systems and introduces the simplest quantity associated with a measurement: effective information, which quantifies its precision, see Proposition \ref{t:classmeas}. Crucially, effective information is \emph{context-dependent} -- it is computed relative to a baseline which may be completely uninformative (the so-called null system) or provided by a subsystem.

Finally entanglement, introduced in \S\ref{s:tangle}, quantifies the obstruction (in bits) to decomposing a measurement into independent submeasurements. It turns out, see discussion after Theorem~\ref{t:g_ei}, that entanglement quantifies the extent to which a measurement is context-dependent -- the extent to which contextual information provided by one submeasurement is useful in understanding another. Theorem~\ref{t:gamma} shows that a measurement is more precise than the sum of its submeasurements \emph{only if} entanglement is non-zero. Precision is thus inextricably bound to context-dependence and indecomposability. The failure of unique descent is thus a feature, not a bug, since it provides ``elbow room'' to build measuring devices that are \emph{not} products of subdevices. 

Space constraints prevent us from providing concrete examples; the interested reader can find these in \cite{bt:08, bt:09, balduzzi:11}. Our running examples are the deterministic set-valued functions
\begin{equation*}
	f:X\rightarrow Y\,\,\,\mbox{ and }\,\,\,g:X\times Y\rightarrow Z
\end{equation*}
which we use to illustrate the concepts as they are developed.

\section{Stochastic maps}
\label{s:stochastic}

Any conditional distribution $p(y|x)$ on finite sets $X$ and $Y$ can be represented as a matrix as follows. Let $\vecify X=\{\varphi:X\rightarrow \bR\}$ denote the vector space of real valued functions on $X$ and similarly for $Y$. $\vecify X$ is equipped with Dirac basis $\{\delta_x:X\rightarrow\bR|x\in X\}$, where
\begin{equation*}
	\delta_x(x') = \left\{\begin{matrix}
		1 & \mbox{if }x=x'\\
		0 & \mbox{else}.
	\end{matrix}\right.
\end{equation*}
Given a conditional distribution $p(y|x)$ construct matrix $\fm_p$ with entry $p(y|x)$ in column $\delta_x$ and row $\delta_y$. Matrix $\fm_p$ is \emph{stochastic}: it has nonnegative entries and its columns sum to 1. Alternatively, given a stochastic matrix $\fm:\vecify X\rightarrow \vecify Y$, we can recover the conditional distribution. The Dirac basis induces Euclidean metric
\begin{equation}
	\bra\bullet|\bullet\ket:\vecify X\otimes\vecify X\rightarrow\bR:
	\left\langle\sum\alpha_x\delta_x\left|\sum\beta_x\delta_x\right\rangle\right.=\sum\alpha_x\beta_x
	\label{e:metric}
\end{equation}
which identifies vector spaces with their duals $\vecify X\approx (\vecify X)^*$. Let $p_\fm(y|x) := \bra\delta_y|\fm(\delta_x)\ket$.

\begin{defn}
	\label{d:catstoch}
	The \emph{category of stochastic maps} $\scat$ has function spaces $\vecify X$ for objects and stochastic matrices $\fm:\vecify X\rightarrow \vecify Y$ with respect to Dirac bases for arrows. We identify of $(\vecify X)^*$ with $\vecify X$ using the Dirac basis without further comment below.
\end{defn}
	
\begin{defn}
	\label{d:dual}
	The \emph{dual} of surjective stochastic map $\fm:\vecify X\rightarrow \vecify Y$ is the composition $\fm^\natural:= \vecify Y\xrightarrow{\fm^*\circ ren} \vecify X$, where $ren$ is the unique map making diagram
	\begin{equation*}
	\xymatrix{
	& (\vecify Y)^*\ar[rr]^{\fm^*} & & 
	(\vecify X)^*\ar[d]_{\omega_X}\\ 
	& (\vecify Y)^*\ar[u]^{ren}\ar[rr]_{\omega_Y}  & & \bR 
	}
	\end{equation*}
	commute. Precomposing $\fm^*$ with $ren$ renormalizes\footnote{If $\fm$ is not surjective, i.e. if one of the rows has all zero entries, then the renormalization is not well-defined.} its columns to sum to 1. The stochastic dual of a stochastic transform is stochastic; further, if $\fm$ is stochastic then $(\fm^\natural)^\natural=\fm$.
\end{defn}

Category $\scat$ is described in terms of braid-like generators and relations in \cite{fritz:09}. A more general, but also more complicated, category of conditional distributions was introduced by Giry \cite{giry:81}, see \cite{panangaden:98}.

\begin{eg}[deterministic functions]
	\label{eg:det}
	Let $\mathtt{FSet}$ be the category of finite sets. Define faithful functor $\vecify:\mathtt{FSet}\rightarrow \scat$ taking set $X$ to $\vecify X$ and function $f:X\rightarrow Y$ to stochastic map $\vecify f:\vecify X\rightarrow \vecify Y:\delta_x\mapsto \delta_{f(x)}$. It is easy to see that $\vecify(X\times Y)=\vecify X\otimes \vecify Y$ and $\vecify (X\cup Y)=\vecify X\times \vecify Y$.

We introduce special notation for commonly used functions:
\begin{itemize}
	\item \emph{Set inclusion.} 
	For any inclusion $i:X\hookrightarrow Y$ of sets, let $\iota:=\vecify i:\vecify X\rightarrow \vecify Y$ denote the corresponding stochastic map. Two important examples are
	\begin{itemize}
		\item \emph{Point inclusion.} 
		Given $x\in X$ define $\iota_x:\bR\rightarrow \vecify X:1\mapsto \delta_x$.
		\item \emph{Diagonal map.} 
		Inclusion $\Delta:X\hookrightarrow X\times X:x\mapsto(x,x)$ induces $\iota_\Delta:\vecify X\rightarrow \vecify X\otimes \vecify X:\delta_x\mapsto \delta_x\otimes \delta_x$.		
	\end{itemize}
	\item \emph{Terminal map.}
	Let $\omega_X:\vecify X\rightarrow \bR:\delta_x\mapsto 1$ denote the terminal map induced by $X\rightarrow\{\bullet\}$.
	\item \emph{Projection.}
	Let $\pi_{XY,X}:\vecify X\otimes \vecify Y\rightarrow \vecify X:\delta_x\otimes \delta_y\mapsto \delta_x$ denote the projection induced by $pr_{X\times Y,X}:X\times Y\rightarrow X:(x,y)\mapsto x$.
\end{itemize}
\end{eg}

\begin{prop}
	[dual is Bayes over uniform distribution]
	\label{t:dual}
	The dual of a stochastic map applies Bayes rule to compute the posterior distribution $\bra \fm^\natural(\delta_y)|\delta_x\ket=p_\fm(x|y)$ using the uniform probability distribution.
\end{prop}

\noindent
Proof: 
The uniform distribution is the dual $\omega_X^\natural:\bR\rightarrow\vecify X:1\mapsto \frac{1}{|X|}\sum_x \delta_x$ of the terminal map $\omega_X:\vecify X\rightarrow \bR$. It assigns equal probability $p_{\omega^\natural}(x)=\frac{1}{|X|}$ to all of $X$'s elements, and can be characterized as the maximally uninformative  distribution \cite{jaynes:57}. Let $\fm:\vecify X\rightarrow \vecify Y$. The normalized transpose is 
\begin{equation*}
	\fm^\natural(\delta_y)= \sum_x \frac{p_\fm(y|x)}{\sum_{x'}p_\fm(y|x')}\delta_x
	= \sum_x \frac{p_\fm(y|x)\cdot p_{\omega^\natural}(x)}{\sum_{x'}p_\fm(y|x')p_{\omega^\natural}(x')}\delta_x
	= \sum_x p_\fm(x|y) \cdot \delta_x.
	\,\,\blacksquare
\end{equation*}

\begin{rem}
	\label{r:notdirac}
	Note that $p_\fm(x|y):=\bra \fm^\natural(\delta_y)|\delta_x\ket\neq\bra \delta_y|\fm(\delta_x)\ket=:p_\fm(y|x)$. Dirac's bra-ket notation must be used with care since stochastic matrices are not necessarily symmetric \cite{dirac:58}.
\end{rem}

\begin{cor}
	[preimages]
	\label{t:preimage}
	The dual $(\vecify f)^\natural:\vecify Y\rightarrow \vecify X$ of stochastic map $\vecify f:\vecify X\rightarrow \vecify Y$ is conditional distribution
	\begin{equation}
		\label{e:preimage}
		p_{\vecify f}(x|y) = \left\{\begin{matrix}
			\frac{1}{|f^{-1}(y)|} & \mbox{if } f(x)=y\\
			0 &\mbox{else}.
		\end{matrix}\right.
	\end{equation}
\end{cor}

\noindent
Proof:
By the proof of Proposition~\ref{t:dual}
\begin{equation*}
		(\vecify f)^\natural(\delta_y)=
		\frac{1}{|f^{-1}(y)|}\sum_{\{x|f(x)=y\}}\delta_x.
		\,\,\blacksquare
\end{equation*}

The support of $p_{\vecify f}(X|y)$ is $f^{-1}(y)$. Elements in the support are assigned equal probability, thereby treating them as an undifferentiated list. Dual $(\vecify f)^\natural$ thus generalizes the inverse image $f^{-1}:Y\rightarrow \underline{2}^X$. Conveniently however, the dual $(\vecify X)^\natural$ simply flips the domain and range of $\vecify f$, whereas the inverse image maps to powerset $\underline{2}^X$, an entirely new object.

\begin{cor}
	[marginalization with respect to uniform distribution]
	\label{t:marginalize}
	Precomposing $\vecify X\otimes \vecify Y\xrightarrow{\fm}\vecify Z$ with the dual $\pi_X^\natural$ to $\vecify X\otimes\vecify Y\xrightarrow{\pi_X}\vecify X$ marginalizes $p_\fm(z|x,y)$ over the uniform distribution on $Y$.
\end{cor}

\noindent
Proof:
By Corollary~\ref{t:preimage} we have $\pi^\natural_X:\vecify X\rightarrow \vecify X\otimes \vecify Y:\delta_y\mapsto \frac{1}{|Y|}\sum_{y\in Y}\delta_x\otimes \delta_y$. It follows immediately that
\begin{equation*}
	p_{\fm\circ\pi^\natural_X}(z|x)=\frac{1}{|Y|}\sum_{y\in Y} p_\fm(z|x,y).
	\,\,\blacksquare
\end{equation*}

Precomposing with $\pi_X^\natural$ treats inputs from $Y$ as extrinsic noise. Although duals can be defined so that they implement Bayes' rule with respect to other probability distributions, this paper restricts attention to the simplest possible renormalization of columns, Definition \ref{d:catstoch}. The uniform distribution is convenient since it uses minimal prior knowledge (it depends only on the number of elements in the set) to generalize pre-images to the stochastic case, Proposition~\ref{t:preimage}.

\section{Distributed dynamical systems}
\label{s:fds}

Probabilistic cellular automata provide useful toy models of a wide range of physical and biological systems. A cellular automaton consists of a collection of cells, each equipped with a mechanism whose output depends on the prior outputs of its neighbors. Two important important examples are

\begin{eg}[Conway's game of life]
	The cellular automaton is a grid of deterministic cells with outputs $\{0,1\}$. A cell outputs 1 at time $t$ iff: (i) three of its neighbors outputted 1s at time $t-1$ or (ii) it and two neighbors outputted 1s at $t-1$. Remarkably, a sufficiently large game of life grid can implement any deterministic computation \cite{berlekamp:82}. 
\end{eg}

\begin{eg}[Hopfield networks]
	These are probabilistic cellular automata \cite{hopfield:82, amit:89}, again with outputs $\{0,1\}$. Cell $n_k$ fires with probability proportional to 
\begin{equation*}
	p(n_{k,t}=1|n_{\bullet,t-1})\propto \exp\left[\frac{1}{T}\sum_{j\rightarrow k}\alpha_{j k}\cdot n_{j,{t-1}}\right].
\end{equation*}
Temperature $T$ controls network stochasticity. Attractors $\{\xi^1,\ldots,\xi^N\}$ are embedded into a network by setting the connectivity matrix as $\alpha_{j k}=\sum_{\mu=1}^N (2\xi_j^\mu-1)(2\xi_k^\mu-1)$.
\end{eg}

It is useful to take a finer perspective on cellular automata by decomposing them into spacetime coordinates or occasions \cite{balduzzi:11}. An occasion $\occ_l=n_{i,t}$ is a cell $n_i$ at a time point $t$. Two occasions are linked $\occ_k\rightarrow \occ_l$ if there is a connection from $\occ_k$'s cell to $\occ_l$'s (because they are neighbors or the same cell) and their time coordinates are $t-1$ and $t$ respectively for some $t$, so occasions form a directed graph. More generally:

\begin{defn}
	\label{d:fds}
	A \emph{distributed dynamical system} $\univ$ consists of the following data:
\end{defn}

\begin{enumerate}[$\univ$1.]
	\item \emph{Directed graph}.
	A graph $G_\univ=(V_\univ,E_\univ)$ with a finite set of vertices or occasions $V_\univ=\{\occ_1\ldots\occ_n\}$ and edges $E_\univ\subset V_\univ\times V_\univ$.
	\item \emph{Alphabets.}
	Each vertex $\occ_l\in V_\univ$ has finite alphabet $\Act_l$ of outputs and finite alphabet $\Sit_l:=\prod_{k\in src(l)}A_k$ of inputs, where $src(l)=\{\occ_k|\occ_k\rightarrow \occ_l\}$.
	\item \emph{Mechanisms.}
	Each vertex $\occ_l$ is equipped with stochastic map $\fm_l:\vecify\Sit_l\rightarrow \vecify\Act_l$.
\end{enumerate}

\begin{figure}[thpb]
	\centering
	\includegraphics[scale=2]{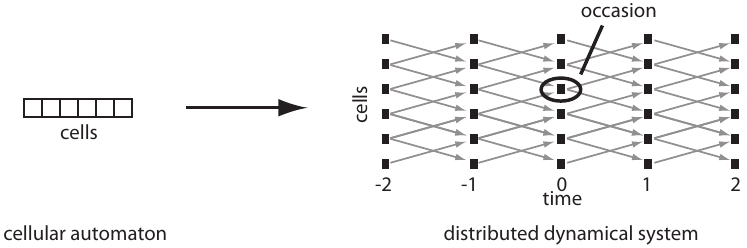}
	\caption{Mapping a cellular automaton to a distributed dynamical system.
	}
	\label{f:dds}
\end{figure}

Taking any cellular automaton over a finite time interval $[t_\alpha,t_\omega]$ initializing the mechanisms at time $t_\alpha$ with fixed values (initial conditions) or probability distributions (noise sources) yields a distributed dynamical system, see Fig.~\ref{f:dds}. Each cell of the original automaton corresponds to a series of occasions in the distributed dynamical system, one per time step.

Cells with memory -- i.e. whose outputs depend on their neighbors outputs over multiple time steps -- receive inputs from occasions more than one time step in the past. If a cell's mechanism changes (learns) over time then different mechanisms are assigned to the cell's occasions at different time points.

The sections below investigate the compositional structure of measurements: how they are built out of submeasurements. Technology for tracking subsystems and submeasurements is therefore necessary. We introduce two closely related categories:

\begin{defn}
	The \emph{category of subsystems} $\subs_\univ$ of $\univ$ is a Boolean lattice with objects given by sets of ordered pairs of vertices $\mangle\in\underline{2}^{V_\univ\times V_\univ}$ and arrows given by inclusions $i_{12}:\mangle_1\hookrightarrow \mangle_2$. The initial and terminal objects are $\bot_\univ=\emptyset$ and $\top_\univ=V_\univ\times V_\univ$.
\end{defn}

\begin{rem}
	\label{r:subsystems}
	Subsystems are defined as ordered pairs of vertices, rather than subgraphs of the directed graph of $\univ$. Pairs of occasions that are not connected by edges are \emph{ineffective}; they do not contribute to the information-processing performed by the system. We include them in the formalism precisely to make their lack of contribution explicit, see Remark~\ref{r:same-point}.
\end{rem}

Let $src(\mangle)=\{\occ_k|(\occ_k,\occ_l)\in\mangle\}$ and similarly for $trg(\mangle)$. Set the input alphabet of $\mangle$ as the product of the output alphabets of its source occasions $\Sit^\mangle=\prod_{src(\mangle)}A_k$ and similarly the output alphabet of $\mangle$ as the product of the output alphabets of its target occasions $\Act^\mangle=\prod_{trg(\mangle)}A_l$. 
\begin{defn}
	The \emph{category of measuring devices} $\mcat_\univ$ on $\univ$ has objects $\Hom_\scat(\vecify \Act^\mangle,\vecify \Sit^\mangle)$ for $\mangle\in \underline{2}^{V_\univ\times V_\univ}$. For $\mangle_1\hookrightarrow \mangle_2$ define arrow 
	\begin{align*}
		r_{21}:\Hom\left(\vecify \Act^{\mangle_2},\vecify \Sit^{\mangle_2}\right) & \rightarrow \Hom\left(\vecify \Act^{\mangle_1},\vecify \Sit^{\mangle_1}\right)\\
		\left[\vecify \Act^{\mangle_2}\xrightarrow{T}\vecify \Sit^{\mangle_2}\right]
		 & \mapsto \left[\vecify \Act^{\mangle_1}\xrightarrow{\pi^\natural_\Act}
		\vecify\Act^{\mangle_2}\xrightarrow{T}\vecify\Sit^{\mangle_2}
		\xrightarrow{\pi_\Sit}\vecify\Sit^{\mangle_1}
		\right],
	\end{align*}
	where $\pi_A$ and $\pi_S$ are shorthands for projections as in Definition \ref{eg:det}.
\end{defn}

The reason for naming $\mcat_\univ$ the category of measuring devices will become clear in \S\ref{s:measurement} below. The two categories are connected by contravariant functor $\psheaf$:

\begin{thm}
	[structure presheaf]
	\label{t:presheaf}
	The \emph{structure presheaf} $\psheaf$ taking
	\begin{equation*}
		\psheaf_\univ:\subs_\univ^{\mathtt{op}}\rightarrow \mcat_\univ:
		\mangle\mapsto\Hom\left(\vecify\Act^\mangle,\vecify\Sit^\mangle\right)
		\mbox{ and }
		i_{12}\mapsto r_{21}
	\end{equation*}
	satisfies the gluing axiom but has non-unique descent.
\end{thm}

\noindent
Proof: Functor $\psheaf$ is trivially a presheaf since it is contravariant. It is an \emph{interesting} presheaf because the gluing axiom holds. 

For gluing we need to show that for any collection $\{\mangle_j\}_{j=1}^l$ of subsystems and sections $\fm_j^\natural \in \psheaf_\univ(\mangle_j)$ such that $r_{j,ji}(\fm_j^\natural)=r_{i,ji}(\fm_i^\natural)$ for all $i$, $j$ there exists section $\fm^\natural\in\psheaf_\univ\left(\bigcup_{j=1}^l\mangle_j\right)$ such that $r_i(\fm^\natural)=\fm^\natural_i$ for all $i$. This reduces to finding a conditional distribution that causes diagram 
\begin{equation*}
	\xymatrix{
	?\ar[r]\ar[d] & \fm_i\ar[d]\\
	\fm_j\ar[r] & \fm_{ij}
	}
\end{equation*}
in $\mcat_\univ$ to commute. The vertices are conditional distributions and the arrows are marginalizations, so rewrite as
\begin{equation*}
	\xymatrix{
	?\ar[r]\ar[d] & p(x,y|u,w)\ar[d]\\
	p(x,z|v,w)\ar[r] & p(x|w),
	}
\end{equation*}
where $p(x|w)=\sum_{v,z}p(x,z|v,w)p^{maxH}(v)$ and similarly for the vertical arrow. It is easy to see that
\begin{equation*}
	p(x,y,z|u,v,w):=\frac{p(x,y|u,w)p(x,z|v,w)}{p(x|w)}
\end{equation*}
satisfies the requirement.

For $\psheaf$ to be a sheaf it would also have to satisfy unique descent: the section satisfying the gluing axiom must not only \emph{exist} for any collection $\{\mangle_j\}_{j=1}^l$ with compatible restrictions but must also be \emph{unique}. Descent in $\psheaf$ is not unique because there are many distributions satisfying the requirement above: strictly speaking $r$ is a marginalization operator rather than restriction. For example, there are many distributions $p(y,z)$ that marginalize to give $p(y)$ and $p(z)$ besides the product distribution $p(y)p(z)$.
\eop

The structure presheaf $\psheaf$ depends on the graph structure and alphabets; mechanisms play no role. We now construct a family of sections of $\psheaf$ using the mechanisms of $\univ$'s occasions. Specifically, given a subsystem $\mangle\in \subs_\univ$, we show how to glue its occasions' mechanisms together to form joint mechanism $\fm_\mangle$. The mechanism $\fm_\univ=\fm_\top$ of the entire system $\univ$ is recovered as a special case. 

In general, subsystem $\mangle$ is not isolated: it receives inputs along edges contained in $\univ$ but \emph{not} in $\mangle$. Inputs along these edges cannot be assigned a fixed value since in general there is no preferred element of $\Act_l$. They also cannot be ignored since $\fm_l$ is defined as receiving inputs from all its sources. Nevertheless, the mechanism of $\mangle$ should depend on $\mangle$ alone. We therefore treat edges not in $\mangle$ as sources of extrinsic noise by marginalizing with respect to the uniform distribution as in Corollary~\ref{t:marginalize}. 

For each vertex $\occ_l\in trg(\mangle)$ let $\Sit^\mangle_l=\prod_{k\in src(l)\cap src(C)} \Act_k$. We then have projection $\pi_l:\vecify \Sit_l\rightarrow \vecify \Sit^\mangle_l$. Define
\begin{equation}
	\label{e:ind_occ}
	\fm^\mangle_l:=\left[\vecify \Sit^\mangle_l\xrightarrow{\pi_l^\natural}\vecify \Sit_l\xrightarrow{\fm_l}\vecify\Act_l\right].
\end{equation}
It follows immediately that $\mangle$ is itself a distributed dynamical system defined by its graph, whose alphabets are inherited from $\univ$ and whose mechanisms are constructed by marginalizing.

Next, we tensor the mechanisms of individual occasions and glue them together using the diagonal map  $\Delta:\Sit^\mangle\rightarrow \prod_{v_l\in trg(\mangle)}\Sit^\mangle_l$. The diagonal map used here\footnote{which is surjective in the sense that all rows contain non-zero entries} generalizes $X\xrightarrow{\Delta}X\times X$ and removes redundancies in $\prod_l \Sit^\mangle_l$, which may, for example, include the same source alphabets many times in different factors.

Let mechanism $\fm_\mangle$ be
\begin{equation}
	\label{e:m_mech}
	\fm_\mangle:=\left[\vecify \Sit^\mangle\xrightarrow{\iota_\Delta}\bigotimes_{\occ_l\in trg(\mangle)}\vecify \Sit^\mangle_l\xrightarrow{\otimes_{\occ_l\in trg(\mangle)}\fm^\mangle_l}\vecify\Act^\mangle\right].
\end{equation}
The dual of $\fm_\mangle$ is 
\begin{equation}
	\label{e:measurement}
	\fm_\mangle^\natural:=\left[\vecify\Act^\mangle\rightarrow \vecify \Sit^\mangle\right].
\end{equation}

Finally, we find that we have constructed a family of sections of $\psheaf$:

\begin{defn}
	\label{d:quale}	
	The \emph{quale} $\mathbf{q}_\univ$ is the family of sections of $\psheaf$ constructed in Eqs.~\eqref{e:ind_occ}, \eqref{e:m_mech} and \eqref{e:measurement}
	\begin{equation*}
		\mathbf{q}_\univ:=\left\{\fm^\natural_\mangle \in \psheaf(\mangle)=
		\Hom\left(\vecify\Act^\mangle,\vecify \Sit^\mangle\right)
		\Big|\mangle\in \subs_\univ\right\}.
	\end{equation*}
\end{defn}
The construction used to glue together the mechanism of the entire system can also be used to construct the mechanism of any subsystem, which provides a window -- the quale -- into the compositional structure of distributed processes.

\section{Measurement}
\label{s:measurement}

This section adapts Definition~\ref{d:cmeasure} to distributed stochastic systems. The first step is to replace elements of state space $X$ with stochastic maps $\uin:\bR\rightarrow \vecify\Sit^\univ$, or equivalently probability distributions on $\Sit^\univ$, which are the system's inputs. Individual elements of $\Sit^\univ$ correspond to Dirac distributions. 

Second, replace function $f:X\rightarrow \bR$ with mechanism $\fm_\univ:\vecify\Sit^\univ\rightarrow \vecify\Act^\univ$. Since we are interested in the compositional structure of measurements we also consider submechanisms $\fm_\mangle$. However, comparing mechanisms requires that they have the same domain and range, so we extend $\fm_\mangle$ to the entire system as follows
\begin{equation}
	\label{e:extension}
	\fm_\mangle = \vecify \Sit^\univ\xrightarrow{\pi}\vecify\Sit^\mangle\xrightarrow{\fm_\mangle}\vecify \Act^\mangle\xrightarrow{\pi^\natural}\vecify \Act^\univ.
\end{equation}

We refer to the extension as $\fm_\mangle$ by abuse of notation. We extend mechanisms implicitly whenever necessary without further comment. Extending mechanisms in this way maps the quale into a cloud of points in $\Hom(\vecify \Act^\univ,\vecify\Sit^\univ)$ labeled by objects in $\subs_\univ$.

In the special case of the initial object $\bot_\univ$, define 
\begin{equation*}
		\label{e:null-extension}
	\fm_\bot = \vecify \Sit^\univ\xrightarrow{\omega}\bR\xrightarrow{\omega^\natural}\vecify \Act^\univ.
\end{equation*}
\begin{rem}
	\label{r:same-point}
	Subsystems differing by non-existent edges (Remark~\ref{r:subsystems}) are mapped to the same mechanism by this construction, thus making the fact that the edges do not exist explicit within the formalism.
\end{rem}

Composing an input with a submechanism yields an output $\uout:=\fm_\mangle\circ \uin: \bR\rightarrow \vecify \Act^\univ$, which is a probability distribution on $\Act^\univ$. We are now in a position to define

\begin{defn}
	\label{d:stochmeas}
	A \emph{measuring device} is the dual $\fm^\natural_\mangle$ to the mechanism of a subsystem. An \emph{output} is a stochastic map $\uout:\bR\rightarrow \vecify\Act^\univ$. A \emph{measurement} is a composition $\fm^\natural_\mangle\circ \uout:\bR\rightarrow \vecify\Sit^\univ$.
\end{defn}


Recall that stochastic maps of the form $\bR\rightarrow \vecify X$ correspond to probability distributions on $X$. Outputs as defined above are thus probability distributions on $\Act^\univ$, the output alphabet of $\univ$. Individual elements of $\Act^\univ$ are recovered as Dirac vectors: $\bR\xrightarrow{\delta_\act}\vecify\Act^\univ$.

\begin{defn}
	\label{d:ei}
	The \emph{effective information} generated by $\mangle_1$ in the \emph{context} of subsystem $\mangle_2\subset \mangle_1$ is
\begin{equation}
	\label{e:rel_ei}
	ei(\fm_{\mangle_2}\rightarrow \fm_{\mangle_1}, \uout) := 
	H\left[\fm_{\mangle_1}^\natural\circ \uout\Big\|\fm_{\mangle_2}^\natural\circ \uout\right].
\end{equation}
The \emph{null context}, corresponding to the empty subsystem $\bot=\emptyset\subset V_\univ\times V_\univ$, is a special case where $\fm_\mangle^\natural\circ \uout$ is replaced by the uniform distribution $\omega_\univ^\natural$ on $\Sit^\univ$. To simplify notation define
\begin{equation*}
	ei(\fm_\mangle,\uout):=ei(\fm_\bot\rightarrow\fm_\mangle,\uout).
\end{equation*}
\end{defn}

Here, $H[p\|q]=\sum_{i}p_i\log_2\frac{p_i}{q_i}$ is the Kullback-Leibler divergence or relative entropy \cite{jaynes:81}. Eq.~\eqref{e:rel_ei} expands as
\begin{equation}
	ei(\fm_{\mangle_2}\rightarrow \fm_{\mangle_1},\uout) 
	 = \sum_{s\in \Sit^\univ} \left\langle\fm^\natural_{\mangle_1}\circ \uout\Big|\delta_s\right\rangle
	\cdot \log_2 \frac{\left\langle\fm^\natural_{\mangle_1}\circ \uout\Big|\delta_s\right\rangle}
	{\left\langle\fm^\natural_{\mangle_2}\circ \uout\Big|\delta_s\right\rangle}.
\end{equation}
When $d_{out}=\delta_\act$ for some $\act\in \Act^\univ$ we have
\begin{equation}
	ei(\fm_{\mangle_2}\rightarrow \fm_{\mangle_1},\delta_\act)
	= \sum_{s\in\Sit^\univ} p_{\fm_{\mangle_1}}(s|\act)\cdot\log_2
	\frac{p_{\fm_{\mangle_1}}(s|\act)}{p_{\fm_{\mangle_2}}(s|\act)}.
\end{equation}

Definition~\ref{d:stochmeas} requires some unpacking. To relate it to the classical notion of measurement, Definition~\ref{d:cmeasure}, we consider system $\univ=\left\{v_X\xrightarrow{f}v_Y\right\}$ where the alphabets of $v_X$ and $v_Y$ are the sets $\Act_{v_X}=X$ and $\Act_{v_Y}=Y$ respectively, and the mechanism of $v_Y$ is $\fm_Y=\vecify f$. In other words, system $\univ$ corresponds to a single deterministic function $f:X\rightarrow Y$.

\begin{prop}
	[classical measurement]
	\label{t:classmeas}
	The measurement $(\vecify f)^\natural\circ \delta_y$ performed when deterministic function $f:X\rightarrow Y$ outputs $y$ is equivalent to the preimage $f^{-1}(y)$. Effective information is $ei(\vecify f,\delta_y)=\log_2\frac{|X|}{|f^{-1}(y)|}$.
\end{prop}

\noindent
Proof:
By Corollary~\ref{t:preimage} measurement $(\vecify f)^\natural\circ \delta_y$ is conditional distribution
\begin{equation*}
	p_{\vecify f}(x|y) = \left\{\begin{matrix}
		\frac{1}{|f^{-1}(y)|} & \mbox{if } f(x)=y\\
		0 &\mbox{else}.
	\end{matrix}\right.
\end{equation*}
which generalizes the preimage. Effective information follows immediately.
\eop

Effective information can be interpreted as quantifying a measurement's precision. It is high if few inputs cause $f$ to output $y$ out of many  -- i.e. $f^{-1}(y)$ has few elements relative to $|X|$  -- and conversely is low if many inputs cause $f$ to output $y$ -- i.e. if the output is relatively insensitive to changes in the input. Precise measurements say a lot about what the input could have been and conversely for vague measurements with low $ei$.

The point of this paper is to develop techniques for studying measurements constructed out of two or more functions. We therefore present computations for the simplest case, distributed system $X\times Y\xrightarrow{g}Z$, in considerable detail. Let $\univ$ be the graph
\begin{equation*}
	\xymatrix{
	v_X\ar[dr] & & v_Y\ar[dl]\\
	& v_Z
	}
\end{equation*}
with obvious assignments of alphabets and the mechanism of $v_Z$ as $\fm_Z=\vecify g$. To make the formulas more readable let $\fm_{XY}=\vecify g$, $\fm_{X\bullet}=\vecify g\circ\pi^\natural_{XY,X}$ and  $\fm_{\bullet Y}=\vecify g\circ\pi^\natural_{XY,Y}$. We then obtain lattice 
\begin{equation}
	\label{e:diag}
	\xymatrix{
		& \top=\fm_{XY} \\
		\fm_{X\bullet}\ar[ur]^{ei(\fm_{X\bullet}\rightarrow\fm_{XY})} & & \fm_{\bullet Y}\ar[ul]_{ei(\fm_{\bullet Y}\rightarrow \fm_{XY})} \\
		& \fm_\bot\ar[ul]^{ei(\fm_{X\bullet})}\ar[ur]_{ei(\fm_{\bullet Y})}\ar[uu]^{ei(\fm_{XY})}
	}
\end{equation}
The remainder of this section and most of the next analyzes measurements in the lattice.

\begin{prop}
	[partial measurement]
	\label{t:confmeas}
	The measurement performed on $X$ when $g:X\times Y\rightarrow Z$  outputs $z$, treating $Y$ as extrinsic noise, is conditional distribution
	\begin{equation}
		\label{e:conf-preimage}
		p(x|z) = \left\{\begin{matrix}
			\frac{|g_{x\times Y}^{-1}(z)|}{|g^{-1}(z)|} &
			\mbox{if } g(x,y)=z\mbox{ for some }y\in Y\\
			0 & \mbox{else,}
		\end{matrix}\right.
	\end{equation}
	where $g^{-1}_{x\times Y}(z):= pr_Y(g^{-1}(z)\cap \{x\}\times Y)$. The effective information generated by the partial measurement is
	\begin{equation}
		\label{e:conf-ei}
		ei\big(\fm_{X\bullet}^\natural,\delta_z\big) 		
		= \log_2|X|+\sum_{x\in X}p(x|z)\cdot
		\log_2 p(x|z).
	\end{equation}
\end{prop}

\noindent
Proof: Treating $Y$ as a source of extrinsic noise yields $\vecify X\xrightarrow{\pi^\natural}\vecify X\otimes \vecify Y\xrightarrow{\vecify g}\vecify Z$ which takes $\delta_x\mapsto \frac{1}{|Y|}\sum_{y\in Y}\delta_{g(x,y)}$. The dual is
\begin{equation*}
	\fm_{X\bullet}^\natural=\pi_{XY,X}\circ (\vecify g)^\natural:\delta_z\mapsto \sum_{x\in X}\frac{|g^{-1}_{x\times Y}(z)|}{|g^{-1}(z)|}\cdot\delta_x.
\end{equation*}	
The computation of effective information follows immediately. 
\eop

A partial measurement is precise if the preimage $g^{-1}(z)$ has small or empty intersection with $\{x\}\times Y$ for most $x$, and large intersection for few $x$.

Propositions~\ref{t:classmeas} and \ref{t:confmeas} compute effective information of a measurement relative to the null context provided by complete ignorance (the uniform distribution). We can also compute the effective information generated by a measurement in the context of a submeasurement:

\begin{prop}
	[relative measurement]
	\label{t:relmeas}
	The information generated by measurement $X\times Y\xrightarrow{g}Z$ in the context of the partial measurement where $Y$ is unobserved noise, is
	\begin{equation}
		\label{e:relmeas}
		ei(\fm_{X\bullet}\rightarrow \fm_{XY},\delta_z)=
		\sum_{x\in X} \frac{g^{-1}_{x\times Y}(z)}{g^{-1}(z)}\log_2
		\frac{|Y|}{g^{-1}_{x\times Y}(z)}.
	\end{equation}
\end{prop}

\noindent
Proof: Applying Propositions~\ref{t:classmeas} and \ref{t:confmeas} obtains
\begin{equation*}
	ei(\fm_{X\bullet}\rightarrow \fm_{XY},\delta_z)=
	\sum_{(x,y)\in g^{-1}(z)}\frac{1}{|g^{-1}(z)|}\log_2\left[
	\frac{1}{|g^{-1}(z)|}\cdot\frac{|g^{-1}(z)|\cdot|Y|}{|g^{-1}_{x\times Y}(z)|}
	\right]
\end{equation*}
which simplifies to the desired expression. 
\eop

To interpret the result decompose $X\times Y\xrightarrow{g} Z$ into a family of functions $\mathcal{R}=\left\{Y\xrightarrow{g_{x\times Y}}Z\big|x\in X\right\}$ labeled by elements of $X$, where $g_{x\times Y}(y):=g(x,y)$. The precision of the measurement performed by $g_{x\times Y}$ is $\log_2\frac{|Y|}{g^{-1}_{x\times Y}(z)}$. It follows that the precision of the relative measurement, Eq.~\eqref{e:relmeas}, is the expected precision of the measurements performed by family $\mathcal{R}$ taken with respect to the probability distribution $p(x|z)=\frac{g^{-1}_{x\times Y}(z)}{g^{-1}(z)}$ generated by the noisy measurement.

In the special case of $g:X\times Y\rightarrow Z$ relative precision is simply the difference of the precision of the larger and smaller subsystems:

\begin{cor}[comparing measurements]
	\label{t:diffmeas}
	\begin{equation*}
		ei(\fm_{X\bullet}\rightarrow \fm_{XY},\delta_z) = ei(\fm_{XY},\delta_z) -ei(\fm_{X\bullet},\delta_z)
	\end{equation*}
\end{cor}

\noindent
Proof: Applying Propositions~\ref{t:classmeas}, \ref{t:confmeas}, \ref{t:relmeas} and simplifying obtains
\begin{align*}
	ei(\fm_{XY},\delta_z) -ei(\fm_{X\bullet},\delta_z) 
	& = \log_2\frac{|X|\cdot|Y|}{|g^{-1}(z)|}
	-\sum_x \frac{|g^{-1}_{x\times Y}(z)|}{|g^{-1}(z)|}\log_2\frac{|X|\cdot |g^{-1}_{x\times Y}(z)|}{|g^{-1}(z)|}\\
	& = \log_2 \frac{|Y|}{|g^{-1}(z)|}+\sum_{(x,y)\in g^{-1}(z)}\frac{1}{|g^{-1}(z)|}
	\log_2 \frac{|g^{-1}(z)|}{|g^{-1}_{x\times Y}(z)|} \\
	& = ei(\fm_{X\bullet}\rightarrow \fm_{XY},\delta_z).
	\,\,\blacksquare
\end{align*}

\section{Entanglement}
\label{s:tangle}

The proof of Theorem~\ref{t:presheaf} showed the structure presheaf has non-unique descent, reflecting the fact that measuring devices do not necessarily reduce to products of subdevices. Similarly, as we will see, measurements do not in general decompose into independent submeasurements. Entanglement, $\gamma$, quantifies how far a measurement diverges in bits from the product of its submeasurements. It turns out that $\gamma>0$ is necessary for a system to generate more information than the sum of its components: non-unique descent thus provides ``room at the top'' to build systems that perform more precise measurements collectively than the sum of their components. 

Entanglement has no direct relation to quantum entanglement. The name was chosen because of a formal resemblance between the two quantities, see Supplementary Information of \cite{bt:09}.

\begin{defn}
	\label{d:gamma}
	\emph{Entanglement} over partition $\cP=\{M_1\ldots M_m\}$ of $src(\fm_\univ)$ is
	\begin{equation*}
		\gamma(\fm_\univ,\cP,\uout)
		=H\left[\fm_\univ^\natural\circ \uout\Big\|\bigotimes_{i=1}^m \pi_j\circ \fm_j^\natural\circ \uout\right]
	\end{equation*}
	where $\pi_j:\vecify \Sit^\univ\rightarrow \vecify \Sit^{M_j}$ and $\fm_j=\{(k,l)\in \fm_\univ|k\in M_j\}$.
\end{defn}

Projecting via $\pi_j$ marginalizes onto the subspace $\vecify\Sit^{M_j}$. Entanglement thus compares the measurement performed by the entire system with submeasurements over the decomposition of the source occasions into partition $\cP$.

\begin{thm}
	[effective information decomposes additively when entanglement is zero]
	\label{t:gamma}
	\begin{equation*}
		\gamma(\fm_\univ,\cP,\uout) = 0
		\,\,\,\,\,\implies\,\,\,\,\,
		ei(\fm_\univ,\uout)=\sum_{i=1}^m ei(\fm_j,\uout).
	\end{equation*}
\end{thm}

\noindent
Proof: Follows from the observations that (i) $H[p\|p_1\otimes p_2]=0$ if and only if $p=p_1\otimes p_2$; (ii) $H[p_1\otimes p_2\| q_1\otimes q_2]=H[p_1\|q_1]+H[p_2\|q_2]$; and (iii) the uniform distribution on $\univ$ is a tensor of uniform distributions on subsystems of $\univ$.
\eop

The theorem shows the relationship between effective information and entanglement. If a system generates more information ``than it should'' (meaning, more than the sum of its subsystems), then the measurements it generates are entangled. Alternatively, only indecomposable measurements can be more precise than the sum of their submeasurements.

We conclude with some detailed computations for $X\times Y\xrightarrow{g}Z$, Diagram~\eqref{e:diag}. Let $\cP=\{X|Y\}$.

\begin{thm}[entanglement and effective information for $g:X\times Y\rightarrow Z$]
	\label{t:g_ei}
	\begin{align*}
		\gamma(\fm_{XY},\cP,\delta_z) & = \sum_{(x,y)\in g^{-1}(z)}
		\frac{1}{|g^{-1}(z)|}\log_2\frac{|g^{-1}(z)|}{|g^{-1}_{x\times Y}(z)|\cdot |g^{-1}_{X\times Y}(z)|} \\
		& = ei(\fm_{XY},\delta_z) - ei(\fm_{X\bullet},\delta_z) - ei(\fm_{\bullet Y},\delta_z).
	\end{align*}
\end{thm}

\noindent
Proof:
The first equality follows from Propositions~\ref{t:classmeas} and \ref{t:confmeas}
\begin{equation*}
	\gamma(\fm_{XY},\cP,\delta_z)=\sum_{(x,y)\in g^{-1}(z)}
	= \sum_{(x,y)\in g^{-1}(z)}\frac{1}{|g^{-1}(z)|}\log_2\left[
	\frac{1}{|g^{-1}(z)|}\cdot \frac{|g^{-1}(z)|}{|g^{-1}_{x\times Y}(z)|}
	\frac{|g^{-1}(z)|}{|g^{-1}_{X\times Y}(z)|}\right].
\end{equation*}
From the same propositions it follows that $ei(\fm_{XY},\delta_z) - ei(\fm_{X\bullet},\delta_z) - ei(\fm_{\bullet Y},\delta_z)$ equals
\begin{gather*}
	 \log_2\frac{|X|\cdot|Y|}{|g^{-1}(x)|}-\sum_{x}\frac{|g^{-1}_{x\times Y}(z)|}{|g^{-1}(z)|}\log_2
	\frac{|X|\cdot|g^{-1}_{x\times Y}(z)|}{|g^{-1}(z)|}
	-\sum_y \frac{|g^{-1}_{X\times y}(z)|}{|g^{-1}(z)|}\log_2\frac{|Y|\cdot|g^{-1}_{X\times y}(z)|}{|g^{-1}(z)|}\\
	= \log_2 \frac{1}{g^{-1}(z)} - \sum_{(x,y)\in g^{-1}(z)}\frac{1}{|g^{-1}(z)|}\cdot
	\log_2\frac{|g^{-1}_{X\times y}(z)|}{|g^{-1}(z)|}\cdot\frac{|g^{-1}_{x\times Y}(z)|}{|g^{-1}(z)|}.
\end{gather*}
Entanglement quantifies how far the size of the pre-image of $g^{-1}(z)$ deviates from the sizes of its $X\times y$ and $x\times Y$ slices as $x$ and $y$ are varied. 
\eop

By Corollary~\ref{t:diffmeas} entanglement also equals $ei(\fm_{X\bullet}\rightarrow \fm_{XY},\delta_z)-ei(\fm_{\bullet Y},\delta_z)$. In Diagram~\eqref{e:diag} entanglement is the vertical arrow minus both arrows at the bottom, or the difference between opposing diagonal arrows. Note that the diagonal arrows from left to right are constructed by adding edge $v_Y\rightarrow v_Z$ to the null system and the subsystem $\fm_{X\bullet}=\{v_X\rightarrow v_Z\}$ respectively.  Entanglement is the difference between the information generated by the diagonal arrows. It quantifies the difference between the information $\{v_Y\rightarrow v_Z\}$ generates in two different contexts.

\begin{cor}
	[characterization of disentangled set-valued functions]
	\label{t:g0z}
	Function $X\times Y\xrightarrow{g}Z$ performs a disentangled measurement when outputting $z$ iff
	\begin{equation*}
		g^{-1}(z)=g^{-1}_{x\times Y}(z)\times g^{-1}_{X\times y}(z)
	\end{equation*}
	for any $x,y$ such that $g(x,y)=z$.
\end{cor}

\noindent
Proof:
By Theorem~\ref{t:g_ei} entanglement is zero iff
\begin{equation*}
	|g^{-1}(z)|=|g^{-1}_{x\times Y}(z)|\cdot |g^{-1}_{X\times y}(z)|
\end{equation*}
for any $x,y$ such that $g(x,y)=z$. This implies the desired result since $g^{-1}(z)\hookrightarrow g^{-1}_{x\times Y}(z)\times g^{-1}_{X\times y}(z)$.
\eop

Thus, the measurement generated by $g$ is disentangled iff its pre-image $g^{-1}(z)$ satisfies a strong geometric ``rectangularity'' constraint: that the pre-image decomposes into the product of its $x\times Y$ and $X\times y$ slices for all pairs of slices intersecting $g^{-1}(z)$. The categorizations performed within a disentangled measuring device have nothing to do with each other, so that the device is best considered as two (or more) distinct devices that happen to have been grouped together for the purposes of performing a computation.

\begin{eg}
	An XOR-gate $g:X\times Y\rightarrow Z$ outputting 0 generates an entangled measurement. The pre-image is $g^{-1}(0)=\{00,11\}$ so the XOR-gate generates 1 bit of information about occasions $v_X$ and $v_Y$. However, the bit is \emph{not localizable}. The measurement generates no information about occasion $v_X$ taken singly, since its output could have been 0 or 1 with equal probability; and similarly for $v_Y$.
\end{eg}

Finally, and unsurprisingly, a function is completely disentangled across all its measurements iff it is a product of two simpler functions:

\begin{cor}
	[completely disentangled functions are products]
	\label{t:g0}
	If $X\times Y\xrightarrow{g}Z$ is surjective, then\\
	$\gamma(\fm_{XY},\cP,\delta_z)=0$ for all $z\in Z$ iff $g$ decomposes into $X\times Y\xrightarrow{g_1\times g_2}Z_1\times Z_2=Z$ for $X\xrightarrow{g_1}Z_1$ and $Y\xrightarrow{g_2}Z_2$.
\end{cor}

\noindent
Proof:
The reverse implication is trivial.
In the forward direction, note that $Z=\{g^{-1}(z)|z\in Z\}$ and, by Corollary \ref{t:g0z}, each pre-image has product structure $g^{-1}(z)=g^{-1}_{x\times Y}(Z)\times g^{-1}_{X\times Y}(z)$. Let $Z_1=\{g^{-1}_{X\times y}|y\in Y\mbox{ and }z\in Z\}$ and similarly for $Z_2$. Define 
\begin{equation*}
	g_1:X\rightarrow Z_1:x\mapsto \mbox{the unique element of form }g^{-1}_{X\times y}(z)
	\mbox{ containing it,}
\end{equation*}
and similarly for $g_2$.
\eop

\section{Discussion}

This paper developed techniques for analyzing the internal structure of distributed measurements. We introduced entanglement, which quantifies the extent to which a measurement is indecomposable. Entanglement can be shown to quantify context-dependence. Moreover, positive entanglement is necessary for a system to generate more information than the sum of its subsystems.  Along the way, we constructed the quale, which geometrically represents the compositional structure of a distributed measurement. The information-theoretic approach developed here is dual, in a precise sense, to the algorithmic perspective on computation. Studying duals $\fm^\natural$ instead of mechanisms $\fm$ shifts the focus from \emph{what} the algorithm does to \emph{how} it does it: instead of analyzing rules we analyze functional dependencies.

The intuition driving the paper is that the structure presheaf $\psheaf$ is an information-theoretic analogue of a tangent space. A particle moving in a manifold $X$ defines a vector field -- a section of the tangent space to $X$, which is a sheaf. The tangent vector at a point depends on the particle's location at ``nearby time-points'': it is computed by taking the limit of difference in positions at $t$ and $t+h$ as $h\rightarrow 0$. Similarly, a system performing a measurement generates a quale, a section of the structure presheaf consisting of ``nearby counterfactuals''. The quale is computed by applying Bayes' rule to determine which inputs could have led to the output.\footnote{A counterfactual input is ``nearby'' to an output if it causes (leads to) that output.} How far this analogy can be developed remains to be seen.

Entanglement can be loosely considered as an information-theoretic analogue of curvature: the extent to which interactions within a system ``warp'' sections of $\psheaf$ away from a product structure. A related approach to geometrically analyzing the complexity of interactions was proposed in \cite{ay:06}. In fact, this project began as an attempt to reformulate \cite{bt:09} in terms of sheaf cohomology using ideas from \cite{ay:06}. We failed at the first step since the structure presheaf is not a sheaf. However, the failure was instructive since it is precisely the \emph{obstruction} to forming a sheaf that is of interest since it is the obstruction (entanglement) that quantifies indecomposability and context-dependence, and only systems whose measurements are entangled are able to generate more information than the sum of their subsystems.

\bibliographystyle{eptcs}
\bibliography{mybib}
\end{document}